\documentclass[italian,english]{article}
\usepackage[T1]{fontenc}
\usepackage[latin1]{inputenc}
\usepackage{color}
\usepackage{graphicx}
\usepackage{amssymb}

\makeatletter

 \newcommand{\lyxaddress}[1]{
   \par {\raggedright #1 
   \vspace{1.4em}
   \noindent\par}
 }

\usepackage{babel}
\makeatother
\begin{document}

\title{\textbf{Extension of the frequency-range of interferometers for the
''magnetic'' components of gravitational waves? }}

\author{\textbf{Christian Corda}}

\maketitle

\lyxaddress{\begin{center}INFN - Sezione di Pisa and Università di Pisa, Via
F. Buonarroti 2, I - 56127 PISA, Italy\end{center}}

\lyxaddress{\begin{center}\textit{E-mail address:} \textcolor{blue}{christian.corda@ego-gw.it} \end{center}}

\begin{abstract}
Recently, with an enlighting treatment, Baskaran and Grishchuk have
shown the presence and importance of the so-called {}``magnetic''
components of gravitational waves (GWs), which have to be taken into
account in the context of the total response functions of interferometers
for GWs propagating from arbitrary directions. In this paper the analysis
of the response functions for the magnetic components is generalized
in its full frequency dependence, while in the work of Baskaran and
Grishchuk the response functions were computed only in the approximation
of wavelength much larger than the linear dimensions of the interferometer.
It is also shown that the response functions to the magnetic components
grow at high frequencies, differently from the values of the response
functions to the well known ordinary components that decrease at high
frequencies. Thus the magnetic components could in principle become
the dominant part of the signal at high frequencies. This is important
for a potential detection of the signal at high frequencies and confirms
that the magnetic contributions must be taken into account in the
data analysis. More, the fact that the response functions of the magnetic
components grow at high frequencies shows that, in principle, the
frequency-range of Earth-based interferometers could extend to frequencies
over 10000 Hz.
\end{abstract}

\lyxaddress{PACS numbers: 04.80.Nn, 04.80.-y, 04.25.Nx}

\section{Introduction}

The design and construction of a number of sensitive detectors for
GWs is underway today. There are some laser interferometers like the
VIRGO detector, being built in Cascina, near Pisa by a joint Italian-French
collaboration \cite{key-1,key-2}, the GEO 600 detector, being built
in Hanover, Germany by a joint Anglo-Germany collaboration \cite{key-3,key-4},
the two LIGO detectors, being built in the United States (one in Hanford,
Washington and the other in Livingston, Louisiana) by a joint Caltech-Mit
collaboration \cite{key-5,key-6}, and the TAMA 300 detector, being
built near Tokyo, Japan \cite{key-7,key-8}. There are many bar detectors
currently in operation too, and several interferometers and bars are
in a phase of planning and proposal stages.

The results of these detectors will have a fundamental impact on astrophysics
and gravitation physics. There will be lots of experimental data to
be analyzed, and theorists will be forced to interact with lots of
experiments and data analysts to extract the physics from the data
stream.

Detectors for GWs will also be important to confirm or ruling out
the physical consistency of General Relativity or of any other theory
of gravitation \cite{key-9,key-10,key-11,key-12}. This is because,
in the context of Extended Theories of Gravity, some differences from
General Relativity and the others theories can be seen starting by
the linearized theory of gravity \cite{key-9,key-10,key-12}. 

With an enlighting treatment, recently, Baskaran and Grishchuk have
shown the presence and importance of the so-called {}``magnetic''
components of GWs, which have to be taken into account in the context
of the total response functions (angular patterns) of interferometers
for GWs propagating from arbitrary directions \cite{key-13}. In this
paper the analysis of the response functions for the magnetic components
is generalized in its full frequency dependence, while in \cite{key-13}
the response functions were computed only in the approximation of
wavelength much larger than the linear dimensions of the interferometer
(i.e. low frequencies). It is also shown that the response functions
to the magnetic components grow at high frequencies, differently from
the values of the response functions to the well known ordinary (called
{}``electric'' in \cite{key-13}) components that decrease at high
frequencies. Thus the magnetic components could in principle become
the dominant part of the signal at high frequencies. This is important
for a potential detection of the signal at high frequencies and confirms
that the magnetic contributions must be taken into account in the
data analysis. More, the fact that the response functions of the magnetic
components grow at high frequencies shows that, in principle, the
frequency-range of Earth-based interferometers could extend to frequencies
over 10000 Hz.

\section{Analysis in the frame of the local observer}

In a laboratory enviroment on earth, the coordinate system in which
the space-time is locally flat is typically used \cite{key-12,key-13,key-14,key-15,key-16}
and the distance between any two points is given simply by the difference
in their coordinates in the sense of Newtonian physics. In this frame,
called the frame of the local observer, GWs manifest themself by exerting
tidal forces on the masses (the mirror and the beam-splitter in the
case of an interferometer, see figure 1). 

\begin{figure}
\includegraphics{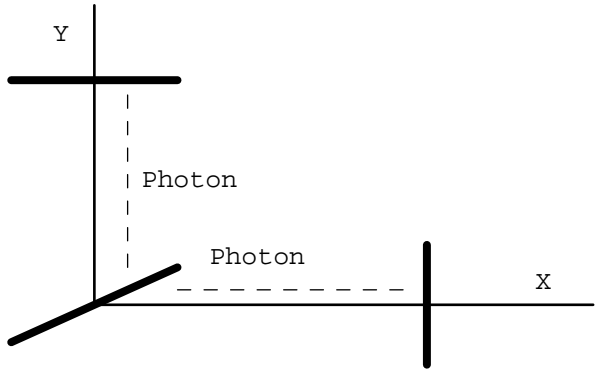}

\caption{photons can be launched from the beam-splitter to be bounced back
by the mirror}
\end{figure}
A detailed analysis of the frame of the local observer is given in
ref. \cite{key-14}, sect. 13.6. Here we remember only the more important
features of this frame:

the time coordinate $x_{0}$ is the proper time of the observer O;

spatial axes are centered in O;

in the special case of zero acceleration and zero rotation the spatial
coordinates $x_{j}$ are the proper distances along the axes and the
frame of the local observer reduces to a local Lorentz frame: in this
case the line element reads 

\begin{equation}
ds^{2}=(-dx^{0})^{2}+\delta_{ij}dx^{i}dx^{j}+O(|dx^{j}|^{2})dx^{\alpha}dx^{\beta};\label{eq: metrica local lorentz}\end{equation}

the effect of GWs on test masses is described by the equation for
geodesic deviation in this frame

\begin{equation}
\ddot{x^{i}}=-\widetilde{R}_{0k0}^{i}x^{k},\label{eq: deviazione geodetiche}\end{equation}
where we have called $\widetilde{R}_{0k0}^{i}$ the linearized Riemann
tensor \cite{key-14}. 

Recently Baskaran and Grishchuk have shown the presence and importance
of the so-called magnetic component of GWs and have computed the detector
pattern in the low frequencies approximation \cite{key-13}. Actually
 this result can be generalized for all GWs (i.e. not only for ones
that have a wavelenght much larger than the arms of the interferometer). 

Before starting with the analysis of the response functions of interferometers,
a brief review of Section 3 of \cite{key-13} is necessary to understand
the importance of the {}``magnetic'' components of GWs. In this
paper we use different notations with respect the ones used in \cite{key-13}.
We work with $G=1$, $c=1$ and $\hbar=1$ and we call $h_{+}(t_{tt}+z_{tt})$
and $h_{\times}(t_{tt}+z_{tt})$ the weak perturbations due to the
$+$ and the $\times$ polarizations of the GW which are expressed
in terms of syncrony coordinates $t_{tt},x_{tt},y_{tt},z_{tt}$ in
the transverse-traceless (TT) gauge. In this way the most general
GW propagating in the $z_{tt}$ direction can be written in terms
of plane monochromatic waves \cite{key-14,key-15,key-16,key-17}

\begin{equation}
\begin{array}{c}
h_{\mu\nu}(t_{tt}+z_{tt})=h_{+}(t_{tt}+z_{tt})e_{\mu\nu}^{(+)}+h_{\times}(t_{tt}+z_{tt})e_{\mu\nu}^{(\times)}=\\
\\=h_{+0}\exp i\omega(t_{tt}+z_{tt})e_{\mu\nu}^{(+)}+h_{\times0}\exp i\omega(t_{tt}+z_{tt})e_{\mu\nu}^{(\times)},\end{array}\label{eq: onda generale}\end{equation}

and the correspondent line element will be

\begin{equation}
ds^{2}=dt_{tt}^{2}-dz_{tt}^{2}-(1+h_{+})dx_{tt}^{2}-(1-h_{+})dy_{tt}^{2}-2h_{\times}dx_{tt}dx_{tt}.\label{eq: metrica TT totale}\end{equation}

The wordlines $x_{tt},y_{tt},z_{tt}=const$ are timelike geodesics
which represent the histories of free test masses \cite{key-14,key-16}.
The coordinate transformation $x^{\alpha}=x^{\alpha}(x_{tt}^{\beta})$
from the TT coordinates to the frame of the local observer is \cite{key-13,key-18}

\begin{equation}
\begin{array}{c}
t=t_{tt}+\frac{1}{4}(x_{tt}^{2}-y_{tt}^{2})\dot{h}_{+}-\frac{1}{2}x_{tt}y_{tt}\dot{h}_{\times}\\
\\x=x_{tt}+\frac{1}{2}x_{tt}h_{+}-\frac{1}{2}y_{tt}h_{\times}+\frac{1}{2}x_{tt}z_{tt}\dot{h}_{+}-\frac{1}{2}y_{tt}z_{tt}\dot{h}_{\times}\\
\\y=y_{tt}+\frac{1}{2}y_{tt}h_{+}-\frac{1}{2}x_{tt}h_{\times}+\frac{1}{2}y_{tt}z_{tt}\dot{h}_{+}-\frac{1}{2}x_{tt}z_{tt}\dot{h}_{\times}\\
\\z=z_{tt}-\frac{1}{4}(x_{tt}^{2}-y_{tt}^{2})\dot{h}_{+}+\frac{1}{2}x_{tt}y_{tt}\dot{h}_{\times}.\end{array}\label{eq: trasf. coord.}\end{equation}

In eqs. (\ref{eq: trasf. coord.}) it is $\dot{h}_{+}\equiv\frac{\partial h_{+}}{\partial t}$
and $\dot{h}_{\times}\equiv\frac{\partial h_{\times}}{\partial t}$.
We emphasize that, in refs. \cite{key-13,key-18} it has been shown
that the linear and quadratics terms, as powers of $x_{tt}^{\alpha}$,
are unambiguously determined by the conditions of the frame of the
local observer.

Considering a free mass riding on a timelike geodesic ($x=l_{1}$,
$y=l_{2},$ $z=l_{3}$) \cite{key-13} eqs. (\ref{eq: trasf. coord.})
define the motion of this mass with respect the introduced frame of
the local observer. It is \begin{equation}
\begin{array}{c}
x(t+z)=l_{1}+\frac{1}{2}[l_{1}h_{+}(t+z)-l_{2}h_{\times}(t+z)]+\frac{1}{2}l_{1}l_{3}\dot{h}_{+}(t+z)+\frac{1}{2}l_{2}l_{3}\dot{h}_{\times}(t+z)\\
\\y(t+z)=l_{2}-\frac{1}{2}[l_{2}h_{+}(t+z)+l_{1}h_{\times}(t+z)]-\frac{1}{2}l_{2}l_{3}\dot{h}_{+}(t+z)+\frac{1}{2}l_{1}l_{3}\dot{h}_{\times}(t+z)\\
\\z(t+z)=l_{3}-\frac{1}{4[}(l_{1}^{2}-l_{2}^{2})\dot{h}_{+}(t+z)+2l_{1}l_{2}\dot{h}_{\times}(t+z),\end{array}\label{eq: Grishuk 0}\end{equation}
which are exactly eqs. (13) of \cite{key-13} rewritten using our
notation. In absence of GWs the position of the mass is $(l_{1},l_{2},l_{3}).$
The effect of the GW is to drive the mass to have oscillations. Thus,
in general, from eqs. (\ref{eq: Grishuk 0}) all three components
of motion are present \cite{key-13}.

Neglecting the terms with $\dot{h}_{+}$and $\dot{h}_{\times}$ in
eqs. (\ref{eq: Grishuk 0}) the {}``traditional'' equations for
the mass motion are obteined \cite{key-14,key-16,key-17}\begin{equation}
\begin{array}{c}
x(t+z)=l_{1}+\frac{1}{2}[l_{1}h_{+}(t+z)-l_{2}h_{\times}(t+z)]\\
\\y(t+z)=l_{2}-\frac{1}{2}[l_{2}h_{+}(t+z)+l_{1}h_{\times}(t+z)]\\
\\z(t+z)=l_{3}.\end{array}\label{eq: traditional}\end{equation}

Cleary, this is the analogue of the electric component of motion in
electrodinamics \cite{key-13}, while equations\begin{equation}
\begin{array}{c}
x(t+z)=l_{1}+\frac{1}{2}l_{1}l_{3}\dot{h}_{+}(t+z)+\frac{1}{2}l_{2}l_{3}\dot{h}_{\times}(t+z)\\
\\y(t+z)=l_{2}-\frac{1}{2}l_{2}l_{3}\dot{h}_{+}(t+z)+\frac{1}{2}l_{1}l_{3}\dot{h}_{\times}(t+z)\\
\\z(t+z)=l_{3}-\frac{1}{4[}(l_{1}^{2}-l_{2}^{2})\dot{h}_{+}(t+z)+2l_{1}l_{2}\dot{h}_{\times}(t+z),\end{array}\label{eq: news}\end{equation}

are the analogue of the magnetic component of motion. One could think
that the presence of these magnetic components is a {}``frame artefact''
due to the transformation (\ref{eq: trasf. coord.}), but it has to
be emphasized that in Section 4 of \cite{key-13} eqs. (\ref{eq: Grishuk 0})
have been obteined directly by the geodesic deviation equation too,
thus the magnetic components have a really physical significance.
The fundamental point of \cite{key-13} is that the magnetic component
becomes important when the frequency of the wave increases, like it
is shown in Section 3 of \cite{key-13}. This can be understood directly
from eqs. (\ref{eq: Grishuk 0}). In fact, using eqs. (\ref{eq: onda generale})
and eqs. (\ref{eq: trasf. coord.}), eqs. (\ref{eq: Grishuk 0}) become\begin{equation}
\begin{array}{c}
x(t+z)=l_{1}+\frac{1}{2}[l_{1}h_{+}(t+z)-l_{2}h_{\times}(t+z)]+\frac{1}{2}l_{1}l_{3}\omega h_{+}(t+z)+\frac{1}{2}l_{2}l_{3}\omega h_{\times}(t+z)\\
\\y(t+z)=l_{2}-\frac{1}{2}[l_{2}h_{+}(t+z)+l_{1}h_{\times}(t+z)]-\frac{1}{2}l_{2}l_{3}\omega h_{+}(t+z)+\frac{1}{2}l_{1}l_{3}\omega h_{\times}(t+z)\\
\\z(t+z)=l_{3}-\frac{1}{4[}(l_{1}^{2}-l_{2}^{2})\omega h_{+}(t+z)+2l_{1}l_{2}\omega h_{\times}(t+z).\end{array}\label{eq: Grishuk 01}\end{equation}

This also means that the terms with $\dot{h}_{+}$ and $\dot{h}_{\times}$
in eqs. (\ref{eq: Grishuk 0}) can be neglectet only when the wavelenght
goes to infinity \cite{key-13}.

To compute the total response functions of interferometers for the
magnetic components generalized in their full frequency dependence,
we will use an analysis parallel to the one used for the first time
in \cite{key-15}: the so called {}``bounching photon metod''. We
emphasize that this metod has been generalized to scalar waves, angular
dependence and massive modes of GWs in \cite{key-12}. We will see
that, in the frame of the local observer, we have to consider two
different effects in the calculation of the variation of the round-trip
time for photons, in analogy with the cases of refs. \cite{key-12,key-15}
where the effects considered were three, but the third effect vanishes
putting the origin of our coordinate system in the beam splitter of
our interferometer (see also the massive case in \cite{key-12}).
Equations (\ref{eq: Grishuk 0}), that represent the coordinates of
the mirror of the interferometer in presence of a GW in the frame
of the local observer, can be rewritten for the pure magnetic component
of the $+$ polarization as

\begin{equation}
\begin{array}{c}
x(t+z)=l_{1}+\frac{1}{2}l_{1}l_{3}\dot{h}_{+}(t+z)\\
\\y(t+z)=l_{2}-\frac{1}{2}l_{2}l_{3}\dot{h}_{+}(t+z)\\
\\z(t+z)=l_{3}-\frac{1}{4}(l_{1}^{2}-l_{2}^{2})\dot{h}_{+}(t+z),\end{array}\label{eq: Grishuk 1}\end{equation}

where $l_{1},l_{2}\textrm{ }and\textrm{ }\textrm{ }l_{3}$ are the
umperturbed coordinates of the mirror. 

To compute the responce functions for an arbitrary incoming direction
of the GW we have to remember that the arms of our interferometer
are in the $\overrightarrow{u}$ and $\overrightarrow{v}$ directions,
while the $x,y,z$ frame is the frame of the local observer in phase
with the frame of the propagating GW. Then we have to make a spatial
rotation of our coordinate system:

\begin{equation}
\begin{array}{ccc}
u & = & -x\cos\theta\cos\phi+y\sin\phi+z\sin\theta\cos\phi\\
\\v & = & -x\cos\theta\sin\phi-y\cos\phi+z\sin\theta\sin\phi\\
\\w & = & x\sin\theta+z\cos\theta,\end{array}\label{eq: rotazione}\end{equation}

or, in terms of the $x,y,z$ frame:

\begin{equation}
\begin{array}{ccc}
x & = & -u\cos\theta\cos\phi-v\cos\theta\sin\phi+w\sin\theta\\
\\y & = & u\sin\phi-v\cos\phi\\
\\z & = & u\sin\theta\cos\phi+v\sin\theta\sin\phi+w\cos\theta.\end{array}\label{eq: rotazione 2}\end{equation}

In this way the GW is propagating from an arbitrary direction $\overrightarrow{r}$
to the interferometer (see figure 2). Because the mirror of eqs. (\ref{eq: Grishuk 1})
is situated in the $u$ direction, using eqs. (\ref{eq: Grishuk 1}),
(\ref{eq: rotazione}) and (\ref{eq: rotazione 2}) the $u$ coordinate
of the mirror is given by

\begin{figure}
\includegraphics{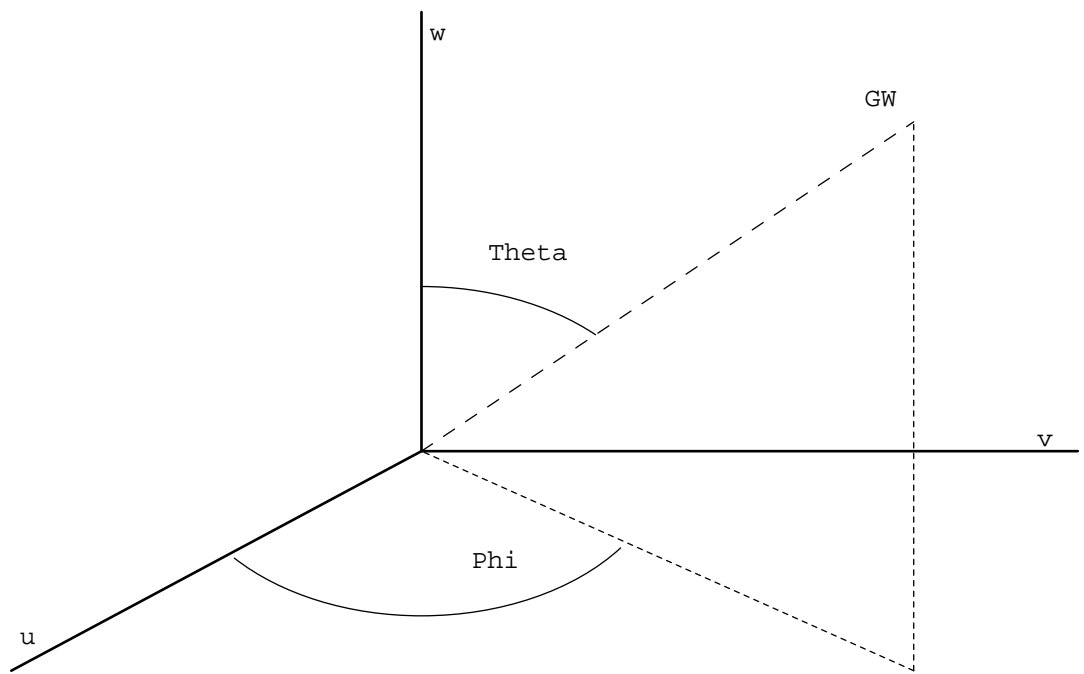}

\caption{a GW incoming from an arbitrary direction}
\end{figure}

\begin{equation}
u=L+\frac{1}{4}L^{2}A\dot{h}_{+}(t+u\sin\theta\cos\phi).\label{eq: du}\end{equation}

where we have defined \begin{equation}
A\equiv\sin\theta\cos\phi(\cos^{2}\theta\cos^{2}\phi-\sin^{2}\phi),\label{eq: A}\end{equation}

and $L=\sqrt{l_{1}^{2}+l_{2}^{2}+l_{3}^{2}}$ is the lenght of the
arms of the interferometer.

A good way to analyze variations in the proper distance (time) is
by means of {}``bouncing photons'' (see refs. \cite{key-12,key-15}
and figure 1). 

We start by considering a photon which propagates in the $u$ axis,
but we will see in Section 4 that the analysis is almost the same
for a photon which propagates in the $v$ axis. 

Putting the origin of our coordinate system in the beam splitter of
our interferometer and using eq. (\ref{eq: du}) the unperturbed coordinates
for the beam-splitter and the mirror are $u_{b}=0$ and $u_{m}=L$.
Thus the unperturbed propagation time between the two masses is

\begin{equation}
T=L.\label{eq: tempo imperturbato}\end{equation}

From eq. (\ref{eq: du}) we find that the displacements of the two
masses under the influence of the GW are

\begin{equation}
\delta u_{b}(t)=0\label{eq: spostamento beam-splitter}\end{equation}

and

\begin{equation}
\delta u_{m}(t)=\frac{1}{4}L^{2}A\dot{h}_{+}(t+L\sin\theta\cos\phi).\label{eq: spostamento mirror}\end{equation}

In this way, the relative displacement, which is defined by

\begin{equation}
\delta L(t)=\delta u_{m}(t)-\delta u_{b}(t)\label{eq: spostamento relativo}\end{equation}

gives

\begin{equation}
\frac{\delta T(t)}{T}=\frac{\delta L(t)}{L}=\frac{1}{4}LA\dot{h}_{+}(t+L\sin\theta\cos\phi).\label{eq: strain magnetico}\end{equation}
But we have the problem that, for a large separation between the test
masses (in the case of Virgo the distance between the beam-splitter
and the mirror is three kilometers, four in the case of LIGO), the
definition (\ref{eq: spostamento relativo}) for relative displacements
becomes unphysical because the two test masses are taken at the same
time and therefore cannot be in a casual connection \cite{key-12,key-15}.
We can write the correct definitions for our bouncing photon like

\begin{equation}
\delta L_{1}(t)=\delta u_{m}(t)-\delta u_{b}(t-T_{1})\label{eq: corretto spostamento B.S. e M.}\end{equation}

and

\begin{equation}
\delta L_{2}(t)=\delta u_{m}(t-T_{2})-\delta u_{b}(t),\label{eq: corretto spostamento B.S. e M. 2}\end{equation}

where $T_{1}$ and $T_{2}$ are the photon propagation times for the
forward and return trip correspondingly. According to the new definitions,
the displacement of one test mass is compared with the displacement
of the other at a later time to allow for finite delay from the light
propagation. We note that the propagation times $T_{1}$ and $T_{2}$
in eqs. (\ref{eq: corretto spostamento B.S. e M.}) and (\ref{eq: corretto spostamento B.S. e M. 2})
can be replaced with the nominal value $T$ because the test mass
displacements are alredy first order in $h_{+}$ \cite{key-12,key-15}.
Thus, for the total change in the distance between the beam splitter
and the mirror in one round-trip of the photon, we get

\begin{equation}
\delta L_{r.t.}(t)=\delta L_{1}(t-T)+\delta L_{2}(t)=2\delta u_{m}(t-T)-\delta u_{b}(t)-\delta u_{b}(t-2T),\label{eq: variazione distanza propria}\end{equation}

and in terms of the amplitude of the GW:

\begin{equation}
\delta L_{r.t.}(t)=\frac{1}{2}L^{2}A\dot{h}_{+}(t+L\sin\theta\cos\phi-L).\label{eq: variazione distanza propria 2}\end{equation}
The change in distance (\ref{eq: variazione distanza propria 2})
lead to changes in the round-trip time for photons propagating between
the beam-splitter and the mirror:

\begin{equation}
\frac{\delta_{1}T(t)}{T}=\frac{1}{2}LA\dot{h}_{+}(t+L\sin\theta\cos\phi-L).\label{eq: variazione tempo proprio 1}\end{equation}

\section{Effect of curved spacetime}

In the last calculation (variations in the photon round-trip time
which come from the motion of the test masses inducted by the magnetic
component of the $+$ polarization of the GW), we implicitly assumed
that the propagation of the photon between the beam-splitter and the
mirror of our interferometer is uniform as if it were moving in a
flat space-time. But the presence of the tidal forces indicates that
the space-time is curved. As a result we have to consider one more
effect after the first discussed that requires spacial separation
\cite{key-12,key-15}. 

From equation (\ref{eq: spostamento mirror}) we get the tidal acceleration
of a test mass caused by the magnetic component of the $+$ polarization
of the GW in the $u$ direction \begin{equation}
\ddot{u}(t+u\sin\theta\cos\phi)=\frac{1}{4}L^{2}A\frac{\partial}{\partial t}\ddot{h}{}_{+}(t+u\sin\theta\cos\phi).\label{eq: acc}\end{equation}

Equivalently we can say that there is a gravitational potential \cite{key-12,key-15}:

\begin{equation}
V(u,t)=-\frac{1}{4}L^{2}A\int_{0}^{u}\frac{\partial}{\partial t}\ddot{h}{}_{+}(t+l\sin\theta\cos\phi)dl,\label{eq:potenziale in gauge Lorentziana}\end{equation}

which generates the tidal forces, and that the motion of the test
mass is governed by the Newtonian equation

\begin{equation}
\ddot{\overrightarrow{r}}=-\bigtriangledown V.\label{eq: Newtoniana}\end{equation}

For the second effect we consider the interval for photons propagating
along the $u$ -axis\begin{equation}
ds^{2}=g_{00}dt^{2}+du^{2}.\label{eq: metrica osservatore locale}\end{equation}

The condition for a null trajectory ($ds=0$) gives the coordinate
velocity of the photons 

\begin{equation}
v^{2}\equiv(\frac{du}{dt})^{2}=1+2V(t,u),\label{eq: velocita' fotone in gauge locale}\end{equation}

which to first order in $h_{+}$ is approximated by

\begin{equation}
v\approx\pm[1+V(t,u)],\label{eq: velocita fotone in gauge locale 2}\end{equation}

with $+$ and $-$ for the forward and return trip respectively. If
we know the coordinate velocity of the photon, we can define the propagation
time for its travelling between the beam-splitter and the mirror:

\begin{equation}
T_{1}(t)=\int_{u_{b}(t-T_{1})}^{u_{m}(t)}\frac{du}{v}\label{eq:  tempo di propagazione andata gauge locale}\end{equation}

and

\begin{equation}
T_{2}(t)=\int_{u_{m}(t-T_{2})}^{u_{b}(t)}\frac{(-du)}{v}.\label{eq:  tempo di propagazione ritorno gauge locale}\end{equation}

The calculations of these integrals would be complicated because the
$u_{m}$ boundaries of them are changing with time:

\begin{equation}
u_{b}(t)=0\label{eq: variazione b.s. in gauge locale}\end{equation}

and

\begin{equation}
u_{m}(t)=L+\delta u_{m}(t).\label{eq: variazione specchio nin gauge locale}\end{equation}

But we note that, to first order in $h{}_{+}$, these contributions
can be approximated by $\delta L_{1}(t)$ and $\delta L_{2}(t)$ (see
eqs. (\ref{eq: corretto spostamento B.S. e M.}) and (\ref{eq: corretto spostamento B.S. e M. 2})).
Thus, the combined effect of the varying boundaries is given by $\delta_{1}T(t)$
in eq. (\ref{eq: variazione tempo proprio 1}). Then we have only
to calculate the times for photon propagation between the fixed boundaries:
$0$ and $L$. We will denote such propagation times with $\Delta T_{1,2}$
to distinguish from $T_{1,2}$. In the forward trip, the propagation
time between the fixed limits is

\begin{equation}
\Delta T_{1}(t)=\int_{0}^{L}\frac{du}{v(t',u)}\approx L-\int_{0}^{L}V(t',u)du,\label{eq:  tempo di propagazione andata  in gauge locale}\end{equation}

where $t'$ is the retardation time (i.e. $t$ is the time at which
the photon arrives in the position $L$, so $L-u=t-t'$) which corresponds
to the unperturbed photon trajectory: 

\begin{center}$t'=t-(L-u)$. \end{center}

Similiary, the propagation time in the return trip is

\begin{equation}
\Delta T_{2}(t)=L-\int_{L}^{0}V(t',u)du,\label{eq:  tempo di propagazione andata  in gauge locale}\end{equation}

where now the retardation time is given by

\begin{center}$t'=t-u$.\end{center}

The sum of $\Delta T_{1}(t-T)$ and $\Delta T_{2}(t)$ give us the
round-trip time for photons traveling between the fixed boundaries.
Then we obtain the deviation of this round-trip time (distance) from
its unperturbed value $2T$ as\begin{equation}
\begin{array}{c}
\delta_{2}T(t)=-\int_{0}^{L}[V(t-2L+u,u)du+\\
\\-\int_{L}^{0}V(t-u,u)]du,\end{array}\label{eq: variazione tempo proprio 2}\end{equation}

and, using eq. (\ref{eq:potenziale in gauge Lorentziana}), it is

\begin{equation}
\begin{array}{c}
\delta_{2}T(t)=\frac{1}{4}L^{2}A\int_{0}^{L}[\int_{0}^{u}\frac{\partial}{\partial t}\ddot{h}_{+}(t-2T+l(1+\sin\theta\cos\phi))dl+\\
\\-\int_{0}^{u}\frac{\partial}{\partial t}\ddot{h}_{+}(t-l(1-\sin\theta\cos\phi)dl]du.\end{array}\label{eq: variazione tempo proprio 2 rispetto h}\end{equation}

Thus we have for the total round-trip proper distance in presence
of the magnetic component of the $+$ polarization of the GW:

\begin{equation}
T_{t}=2T+\delta_{1}T+\delta_{2}T,\label{eq: round-trip  totale in gauge locale}\end{equation}

and\begin{equation}
\delta T_{u}=T_{t}-2T=\delta_{1}T+\delta_{2}T\label{eq:variaz round-trip totale in gauge locale}\end{equation}

is the total variation of the proper time (distance) for the round-trip
of the photon in presence of the GW in the $u$ direction.

Using eqs. (\ref{eq: variazione tempo proprio 1}), (\ref{eq: variazione tempo proprio 2 rispetto h})
and the Fourier transform of $h_{+}$ defined by

\begin{equation}
\tilde{h}_{+}(\omega)=\int_{-\infty}^{\infty}dth_{+}(t)\exp(i\omega t),\label{eq: trasformata di fourier}\end{equation}

the quantity (\ref{eq:variaz round-trip totale in gauge locale})
can be computed in the frequency domain as 

\begin{equation}
\tilde{\delta}T_{u}(\omega)=\tilde{\delta}_{1}T(\omega)+\tilde{\delta}_{2}T(\omega)\label{eq:variaz round-trip totale in gauge locale 2}\end{equation}

where

\begin{equation}
\tilde{\delta}_{1}T(\omega)=-i\omega\exp[i\omega L(1-\sin\theta\cos\phi)]\frac{L^{2}A}{2}\tilde{h}_{+}(\omega)\label{eq: dt 1 omega}\end{equation}

\begin{equation}
\begin{array}{c}
\tilde{\delta}_{2}T(\omega)=\frac{i\omega L^{2}A}{4}[\frac{-1+\exp[i\omega L(1-\sin\theta\cos\phi)]-iL\omega(1-\sin\theta\cos\phi)}{(1-\sin\theta\cos\phi)^{2}}+\\
\\+\frac{\exp(2i\omega L)(1-\exp[i\omega L(-1-\sin\theta\cos\phi)]-iL\omega(1+\sin\theta\cos\phi)}{(-1-\sin\theta\cos\phi)^{2}}]\tilde{h}_{+}(\omega).\end{array}\label{eq: dt 2 omega}\end{equation}

In the above computation the derivation and translation theorems of
the Fourier transform have been used. In this way we obtain the response
function of the $u$ arm of our interferometer to the magnetic component
of the $+$ polarization of the GW as

\begin{equation}
\begin{array}{c}
H_{u}^{+}(\omega)\equiv\frac{\tilde{\delta}T_{u}(\omega)}{L\tilde{h}_{+}(\omega)}=\\
\\=-i\omega\exp[i\omega L(1-\sin\theta\cos\phi)]\frac{LA}{2}+\\
\\\frac{i\omega LA}{4}[\frac{-1+\exp[i\omega L(1-\sin\theta\cos\phi)]-iL\omega(1-\sin\theta\cos\phi)}{(1-\sin\theta\cos\phi)^{2}}+\\
\\+\frac{\exp(2i\omega L)(1-\exp[i\omega L(-1-\sin\theta\cos\phi)]-iL\omega(1+\sin\theta\cos\phi)}{(-1-\sin\theta\cos\phi)^{2}}].\end{array}\label{eq: risposta u}\end{equation}

\section{Computation for the $v$ arm}

The computation for the $v$ arm is parallel to the one above. Using
eqs. (\ref{eq: Grishuk 1}), (\ref{eq: rotazione}) and (\ref{eq: rotazione 2})
the coordinate of the mirror in the $v$ arm is:

\begin{equation}
v=L+\frac{1}{4}L^{2}B\dot{h}_{+}(t+v\sin\theta\sin\phi),\label{eq: dv}\end{equation}

where we have defined \begin{equation}
B\equiv\sin\theta\sin\phi(\cos^{2}\theta\cos^{2}\phi-\sin^{2}\phi).\label{eq: B}\end{equation}

Thus, with the same way of thinking of previous Sections, we get variations
in the photon round-trip time which come from the motion of the beam-splitter
and the mirror in the $v$ direction:

\begin{equation}
\frac{\delta_{1}T(t)}{T}=\frac{1}{2}LB\dot{h}_{+}(t+L\sin\theta\sin\phi-L),\label{eq: variazione tempo proprio 1 in v}\end{equation}

while the second contribute (propagation in a curve spacetime) will
be 

\begin{equation}
\begin{array}{c}
\delta_{2}T(t)=\frac{1}{4}L^{2}B\int_{0}^{L}[\int_{0}^{u}\frac{\partial}{\partial t}\ddot{h}_{+}(t-2T+l(1-\sin\theta\sin\phi))dl+\\
\\-\int_{0}^{u}\frac{\partial}{\partial t}\ddot{h}_{+}(t-l(1-\sin\theta\sin\phi)dl]du,\end{array}\label{eq: variazione tempo proprio 2 rispetto h in v}\end{equation}

and the total response function of the $v$ arm for the magnetic component
of the $+$ polarization of GWs is given by\begin{equation}
\begin{array}{c}
H_{v}^{+}(\omega)\equiv\frac{\tilde{\delta}T_{u}(\omega)}{L\tilde{h}_{+}(\omega)}=\\
\\=-i\omega\exp[i\omega L(1-\sin\theta\sin\phi)]\frac{LB}{2}+\\
\\+\frac{i\omega LB}{4}[\frac{-1+\exp[i\omega L(1-\sin\theta\sin\phi)]-iL\omega(1-\sin\theta\sin\phi)}{(1-\sin\theta\cos\phi)^{2}}+\\
\\+\frac{\exp(2i\omega L)(1-\exp[i\omega L(-1-\sin\theta\sin\phi)]-iL\omega(1+\sin\theta\sin\phi)}{(-1-\sin\theta\sin\phi)^{2}}].\end{array}\label{eq: risposta v}\end{equation}

\section{The total response function of an interferometer for the $+$ polarization}

The total response function for the magnetic component of the $+$
polarization is given by the difference of the two response function
of the two arms:\begin{equation}
H_{tot}^{+}(\omega)\equiv H_{u}^{+}(\omega)-H_{v}^{+}(\omega),\label{eq: risposta totale}\end{equation}

and using eqs. (\ref{eq: risposta u}) and (\ref{eq: risposta v})
we obtain a complicated formula\begin{equation}
\begin{array}{c}
H_{tot}^{+}(\omega)=\frac{\tilde{\delta}T_{tot}(\omega)}{L\tilde{h}_{+}(\omega)}=\\
\\=-i\omega\exp[i\omega L(1-\sin\theta\cos\phi)]\frac{LA}{2}+\frac{LB}{2}i\omega\exp[i\omega L(1-\sin\theta\sin\phi)]\\
\\-\frac{i\omega LA}{4}[\frac{-1+\exp[i\omega L(1-\sin\theta\cos\phi)]-iL\omega(1-\sin\theta\cos\phi)}{(1-\sin\theta\cos\phi)^{2}}\\
\\+\frac{\exp(2i\omega L)(1-\exp[i\omega L(-1-\sin\theta\cos\phi)]-iL\omega(1+\sin\theta\cos\phi)}{(-1-\sin\theta\cos\phi)^{2}}]+\\
\\+\frac{i\omega LB}{4}[\frac{-1+\exp[i\omega L(1-\sin\theta\sin\phi)]-iL\omega(1-\sin\theta\sin\phi)}{(1-\sin\theta\cos\phi)^{2}}+\\
\\+\frac{\exp(2i\omega L)(1-\exp[i\omega L(-1-\sin\theta\sin\phi)]-iL\omega(1+\sin\theta\sin\phi)}{(-1-\sin\theta\sin\phi)^{2}}],\end{array}\label{eq: risposta totale 2}\end{equation}

that, in the low freuencies limit is in perfect agreement with the
result of Baskaran and Grishchuk (eq. 49 of \cite{key-13}): \begin{equation}
H_{tot}^{+}(\omega\rightarrow0)=\frac{1}{4}\sin\theta[(\cos^{2}\theta+\sin2\phi\frac{1+\cos^{2}\theta}{2})](\cos\phi-\sin\phi).\label{eq: risposta totale bassa}\end{equation}

We emphasize that in our work the $x,y,z$ frame is the frame of the
local observer in phase with respect the propagating GW, while in
\cite{key-13} the two frames are not in phase (i.e. in our work the
third angle is put equal to zero, this is not a restriction as it
is known in literature, see for example \cite{key-12}).

In figures 3 and 4 the absolute value of the total response functions
(\ref{eq: risposta totale 2}) of the Virgo and LIGO interferometers
to the magnetic component of the $+$ polarization of GWs for $\theta=\frac{\pi}{4}$
and $\phi=\frac{\pi}{3}$ are respectively shown. This value grows
at high frequencies.%
\begin{figure}
\includegraphics{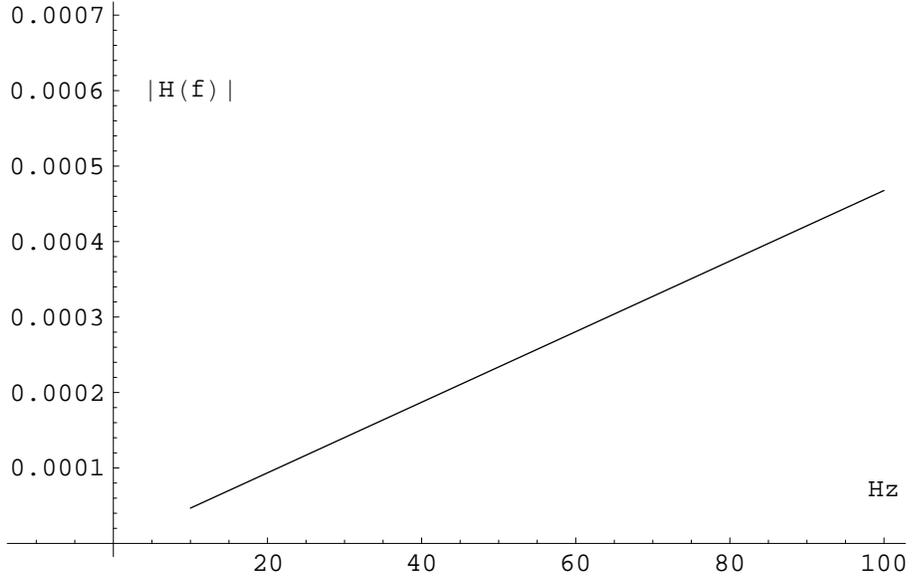}

\caption{the absolute value of the total response function of the Virgo interferometer
to the magnetic component of the $+$ polarization of GWs for $\theta=\frac{\pi}{4}$
and $\phi=\frac{\pi}{3}$}
\end{figure}
\begin{figure}
\includegraphics{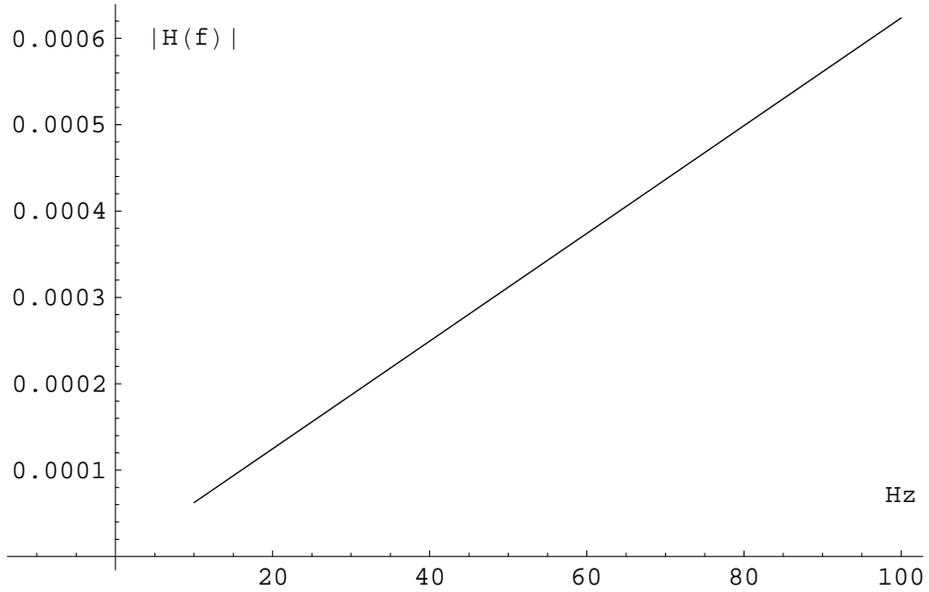}

\caption{the absolute value of the total response function of the LIGO interferometer
to the magnetic component of the $+$ polarization of GWs for $\theta=\frac{\pi}{4}$
and $\phi=\frac{\pi}{3}$}
\end{figure}
In figure 5 the angular dependence of the total response function
(\ref{eq: risposta totale 2}) of the Virgo and LIGO interferometers
to the magnetic component of the $+$ polarization of GWs for $f=100Hz$
is shown. 

\begin{figure}
\includegraphics{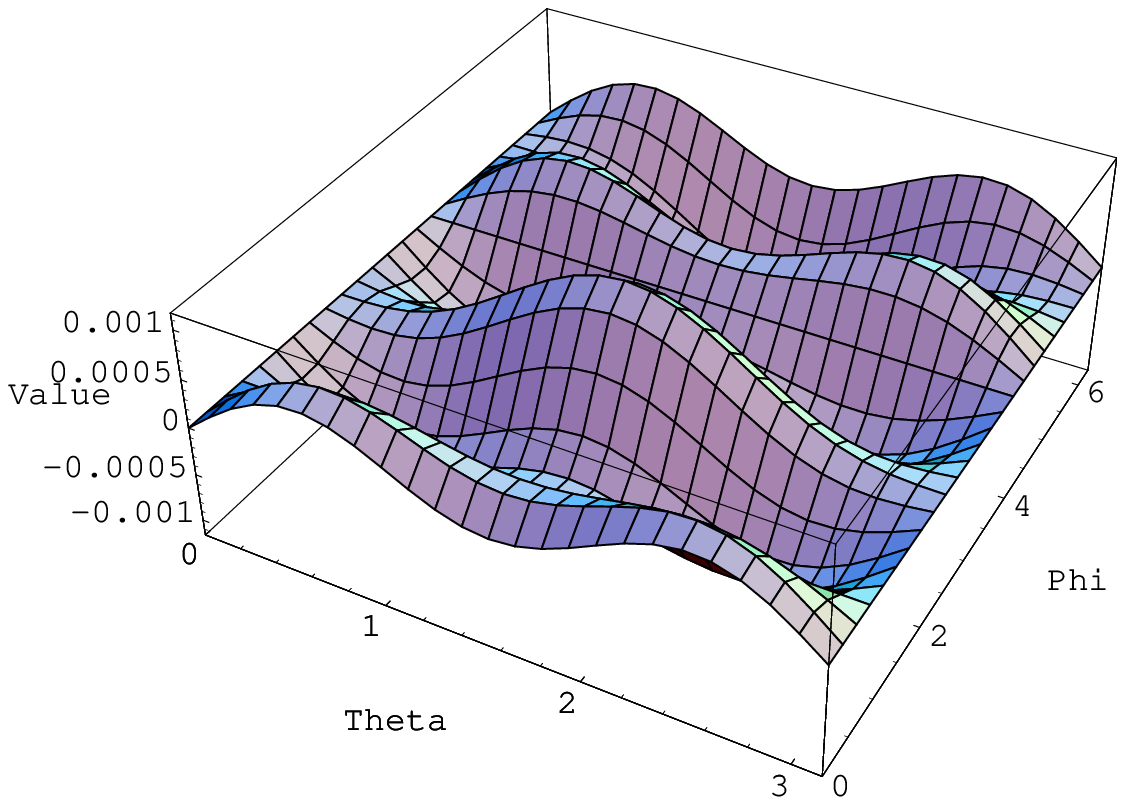}

\caption{the angular dependence of the total response function of the Virgo
and LIGO interferometers to the magnetic component of the $+$ polarization
of GWs for $f=100Hz$}
\end{figure}

\section{Analysis for the $\times$ polarization}

The analysis can be generalized for the magnetic component of the
$\times$ polarization too. In this case, equations (\ref{eq: Grishuk 0})
can be rewritten for the pure magnetic component of the $\times$
polarization as

\begin{equation}
\begin{array}{c}
x(t+z)=l_{1}+\frac{1}{2}l_{2}l_{3}\dot{h}_{\times}(t+z)\\
\\y(t+z)=l_{2}+\frac{1}{2}l_{1}l_{3}\dot{h}_{\times}(t+z)\\
\\z(t+z)=l_{3}-\frac{1}{2}l_{1}l_{2}\dot{h}_{\times}(t+z).\end{array}\label{eq: Grishuk 2}\end{equation}

Using eqs. (\ref{eq: Grishuk 2}), (\ref{eq: rotazione}) and (\ref{eq: rotazione 2})
the $u$ coordinate of the mirror situated in the $u$ arm of the
interferometer is given by \begin{equation}
u=L+\frac{1}{4}L^{2}C\dot{h}_{\times}(t+u\sin\theta\cos\phi).\label{eq: du C}\end{equation}

where we have defined \begin{equation}
C\equiv-2\cos\theta\cos^{2}\phi\sin\theta\sin\phi,\label{eq: C}\end{equation}
while the $v$ coordinate of the mirror situated in the $v$ arm of
the interferometer is given by \begin{equation}
v=L+\frac{1}{4}L^{2}D\dot{h}_{\times}(t+v\sin\theta\sin\phi).\label{eq: dv  D}\end{equation}

where it is\begin{equation}
D\equiv2\cos\theta\cos\phi\sin\theta\sin^{2}\phi.\label{eq: D}\end{equation}

Thus, with an analysis parallel to the one of previous Sections, it
is possible to show that the total response function of the interferometer
for the magnetic component of the $\times$ polarization of GWs is\begin{equation}
\begin{array}{c}
H_{tot}^{\times}(\omega)=\frac{\tilde{\delta}T_{tot}(\omega)}{L\tilde{h}_{\times}(\omega)}=\\
\\=-i\omega\exp[i\omega L(1-\sin\theta\cos\phi)]\frac{LC}{2}+\frac{LD}{2}i\omega\exp[i\omega L(1-\sin\theta\sin\phi)]\\
\\-\frac{i\omega LC}{4}[\frac{-1+\exp[i\omega L(1-\sin\theta\cos\phi)]-iL\omega(1-\sin\theta\cos\phi)}{(1-\sin\theta\cos\phi)^{2}}\\
\\+\frac{\exp(2i\omega L)(1-\exp[i\omega L(-1-\sin\theta\cos\phi)]-iL\omega(1+\sin\theta\cos\phi)}{(-1-\sin\theta\cos\phi)^{2}}]+\\
\\+\frac{i\omega LD}{4}[\frac{-1+\exp[i\omega L(1-\sin\theta\sin\phi)]-iL\omega(1-\sin\theta\sin\phi)}{(1-\sin\theta\cos\phi)^{2}}+\\
\\+\frac{\exp(2i\omega L)(1-\exp[i\omega L(-1-\sin\theta\sin\phi)]-iL\omega(1+\sin\theta\sin\phi)}{(-1-\sin\theta\sin\phi)^{2}}],\end{array}\label{eq: risposta totale 2 per}\end{equation}

that, in the low freuencies limit is in perfect agreement with the
result of Baskaran and Grishchuk (eq. 50 of \cite{key-13}): \begin{equation}
H_{tot}^{\times}(\omega\rightarrow0)=\frac{1}{4}\sin2\phi(\cos\phi+\sin\phi)\cos\theta.\label{eq: risposta totale bassa per}\end{equation}
In figure 6 and 7 the absolute value of the total response functions
(\ref{eq: risposta totale 2 per}) of the Virgo and LIGO interferometers
to the magnetic component of the $\times$ polarization of GWs for
$\theta=\frac{\pi}{4}$ and $\phi=\frac{\pi}{3}$ are respectively
shown. This value grows at high frequencies in analogy with the case
seen in previous Section for the magnetic component of the $+$ polarization.
\begin{figure}
\includegraphics{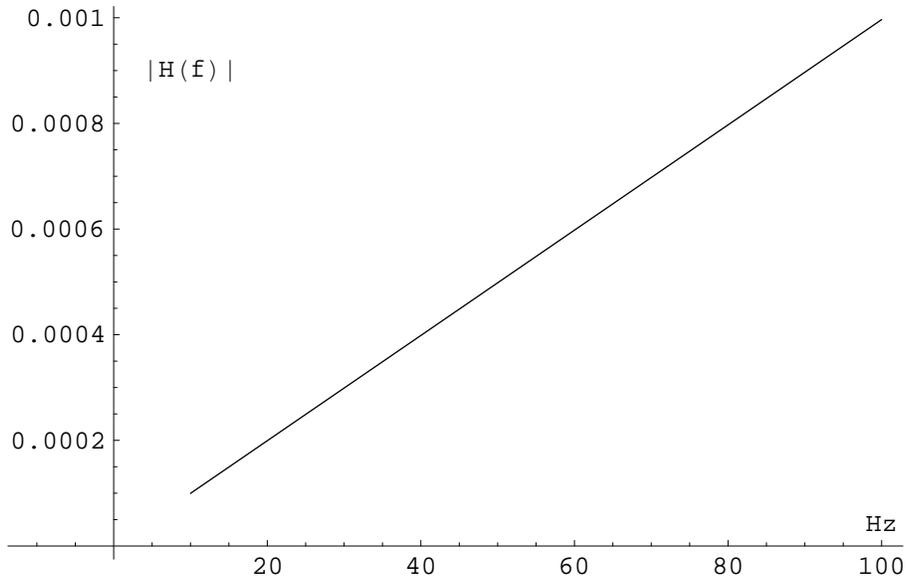}

\caption{the absolute value of the total response function of the Virgo interferometer
to the magnetic component of the $\times$ polarization of GWs for
$\theta=\frac{\pi}{4}$ and $\phi=\frac{\pi}{3}$}
\end{figure}
\begin{figure}
\includegraphics{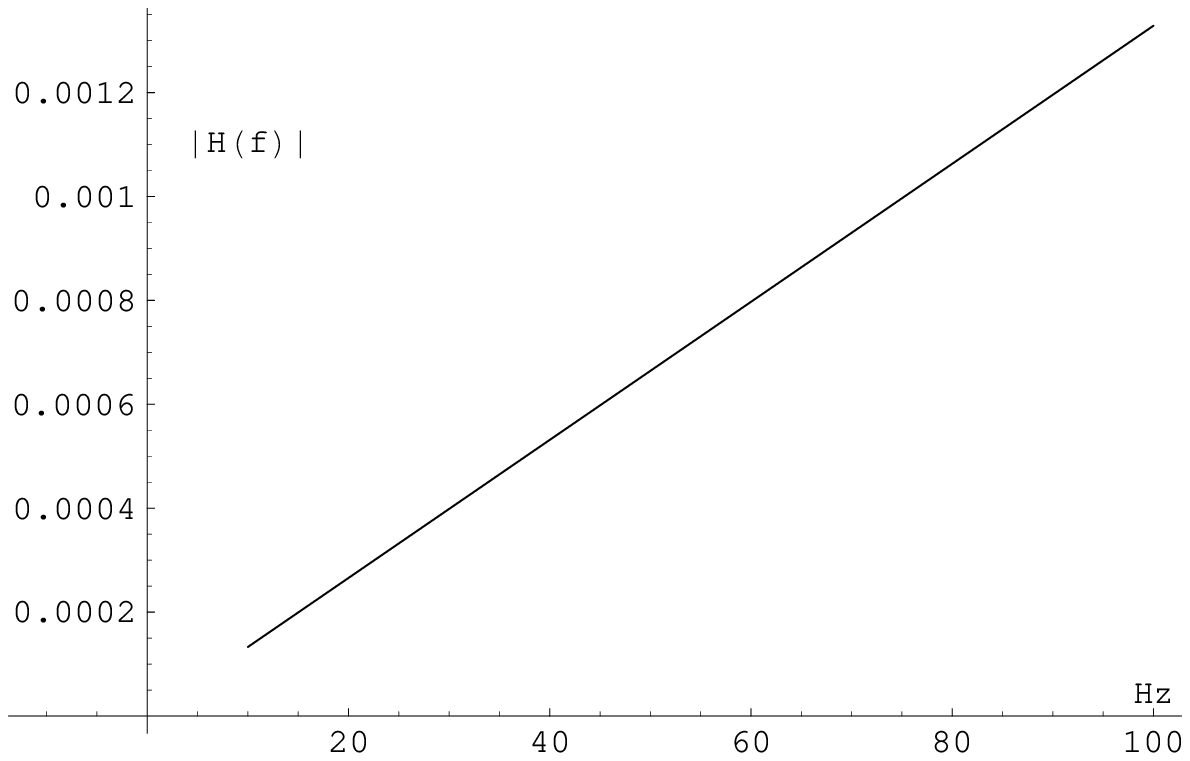}

\caption{the absolute value of the total response function of the LIGO interferometer
to the magnetic component of the $\times$ polarization of GWs for
$\theta=\frac{\pi}{4}$ and $\phi=\frac{\pi}{3}$}
\end{figure}
In figure 8 the angular dependence of the total response function
(\ref{eq: risposta totale 2 per}) of the Virgo and LIGO interferometers
to the magnetic component of the $\times$ polarization of GWs for
$f=100Hz$ is shown. %
\begin{figure}
\includegraphics{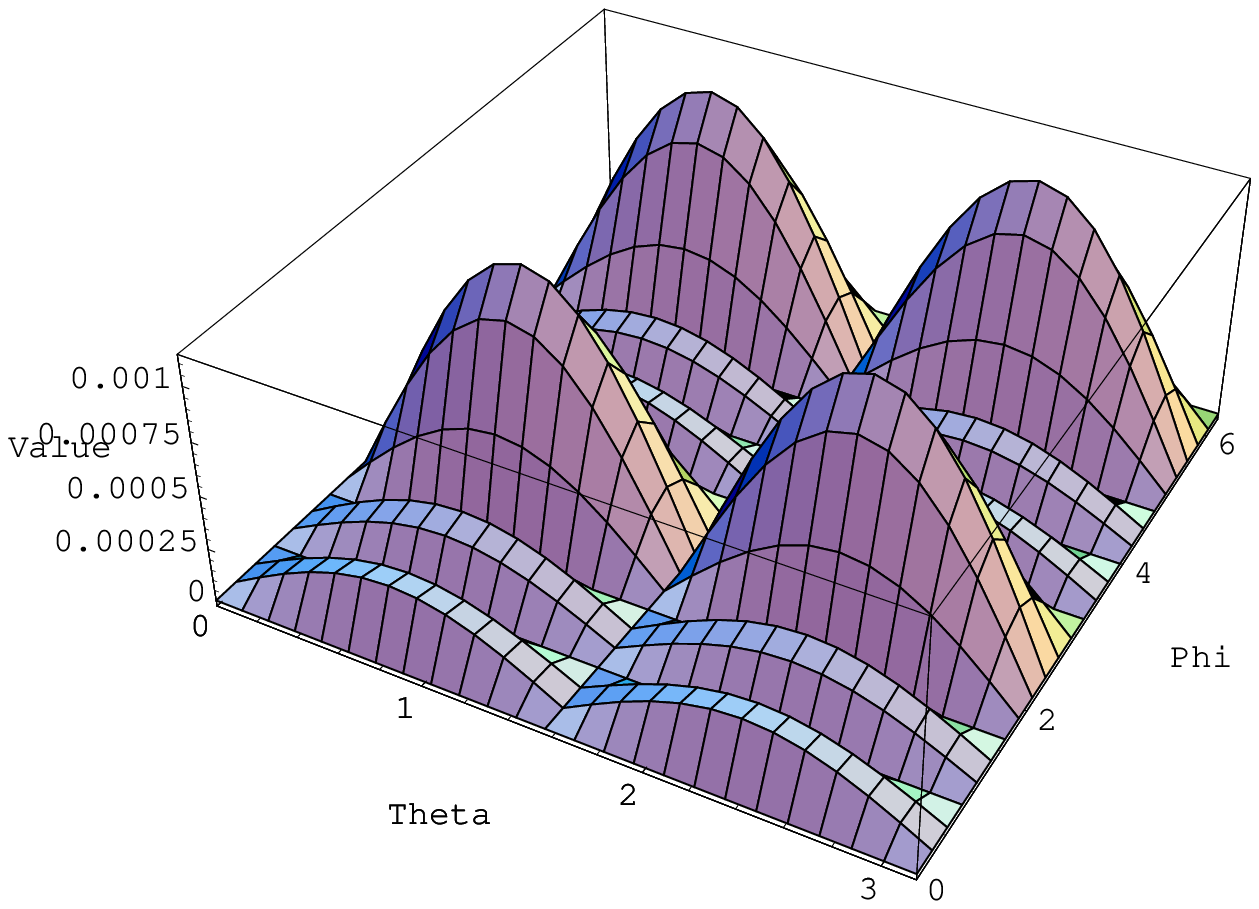}

\caption{the angular dependence of the total response function of the Virgo
and LIGO interferometers to the magnetic component of the $\times$
polarization of GWs for $f=100Hz$}
\end{figure}

\section{Extension of the frequency-range for Earth-based interferometers?}

The fact that the values of the response functions of interferometers
to the {}``electric'' components (i.e. the well known ordinary components)
decrease with frequency is well known in literature. This is because
the response functions to the electric components are proportional
to the function

\begin{equation}
sinc(\omega L)\equiv\frac{\sin\omega L}{\omega L},\label{eq: sinc}\end{equation}
see for example equation 12 of \cite{key-15} and equation 2.23 of
\cite{key-19}. Then, because the response functions to the {}``magnetic''
components grow with frequency, as it is shown in eqs. (\ref{eq: risposta totale 2})
and (\ref{eq: risposta totale 2 per}) and in their graphics of figures
3,4, 6 and 7, the part of signal which arises from the magnetic components
could in principle become the dominant part of the signal at high
frequencies (the fact that the response functions to the magnetic
components grow with frequency is due to eqs. (\ref{eq: Grishuk 01})
which show that the motion of test masses is proportional to the frequency).
This is important for a potential detection of the signal at high
frequencies and confirms the result of \cite{key-13} that the magnetic
contributions must be taken into account in the data analysis in addition
of the response functions of the electric components which are well
known in literature (see for example \cite{key-19,key-20,key-21,key-22}).
The fact that the response functions of the magnetic components grow
at high frequencies also shows that, in principle, the frequency-range
of Earth-based interferometers could extend to frequencies over 10000
Hz. In figures 9,10 and 11,12 the absolute value of the total response
functions (\ref{eq: risposta totale 2}) and (\ref{eq: risposta totale 2 per})
of the Virgo and LIGO interferometers to the magnetic components of
the $+$ and $\times$ polarization of GWs for $\theta=\frac{\pi}{4}$
and $\phi=\frac{\pi}{3}$ are extended to the range $10000Hz\leq f\leq20000Hz$.
\begin{figure}
\includegraphics{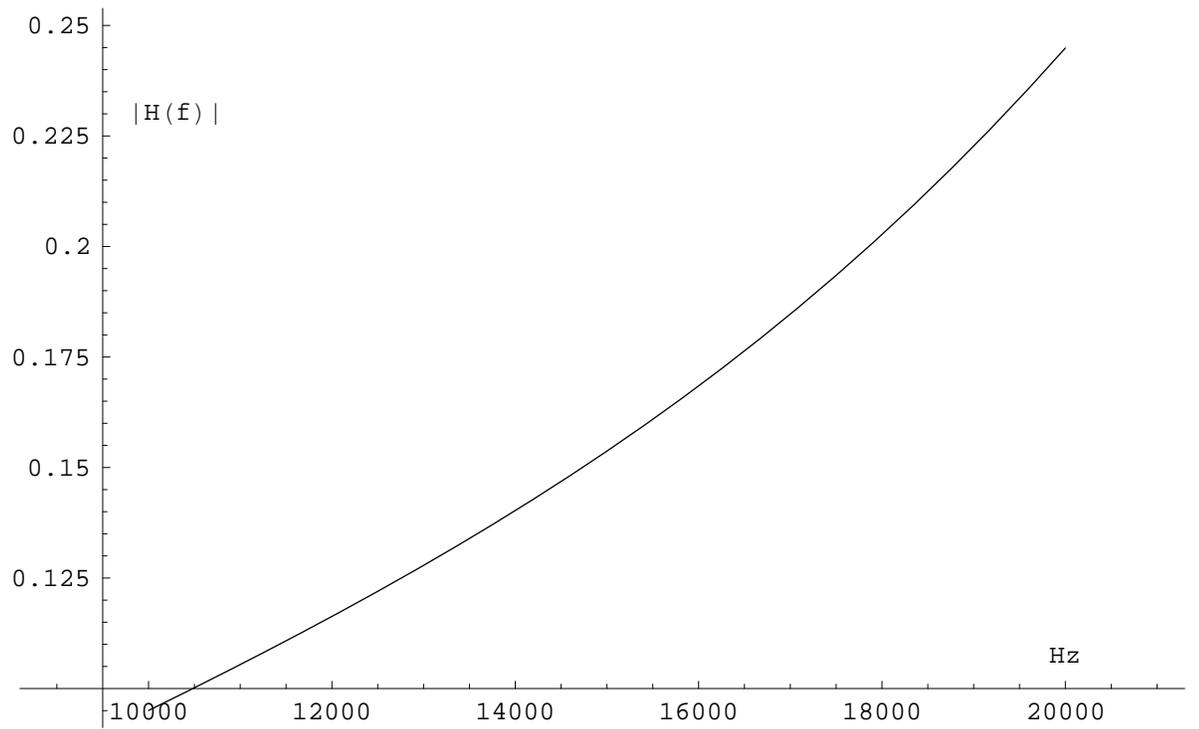}

\caption{the absolute value of the total response function of the Virgo interferometer
to the magnetic component of the $+$ polarization of GWs for $\theta=\frac{\pi}{4}$
and $\phi=\frac{\pi}{3}$ in the frequency-range $10000Hz\leq f\leq20000Hz$}
\end{figure}
\begin{figure}
\includegraphics{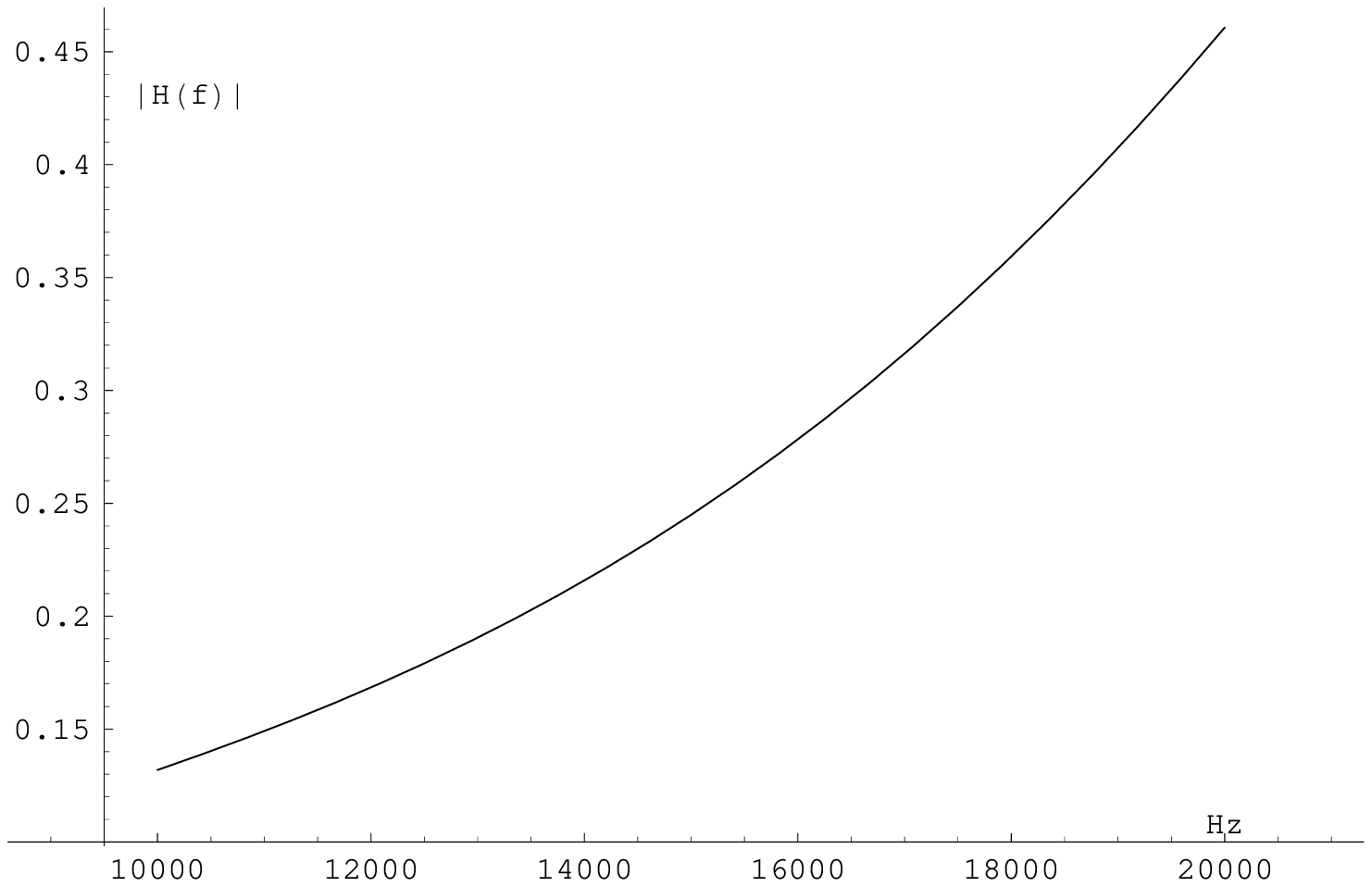}

\caption{the absolute value of the total response function of the LIGO interferometer
to the magnetic component of the $+$ polarization of GWs for $\theta=\frac{\pi}{4}$
and $\phi=\frac{\pi}{3}$ in the frequency-range $10000Hz\leq f\leq20000Hz$}
\end{figure}
\begin{figure}
\includegraphics{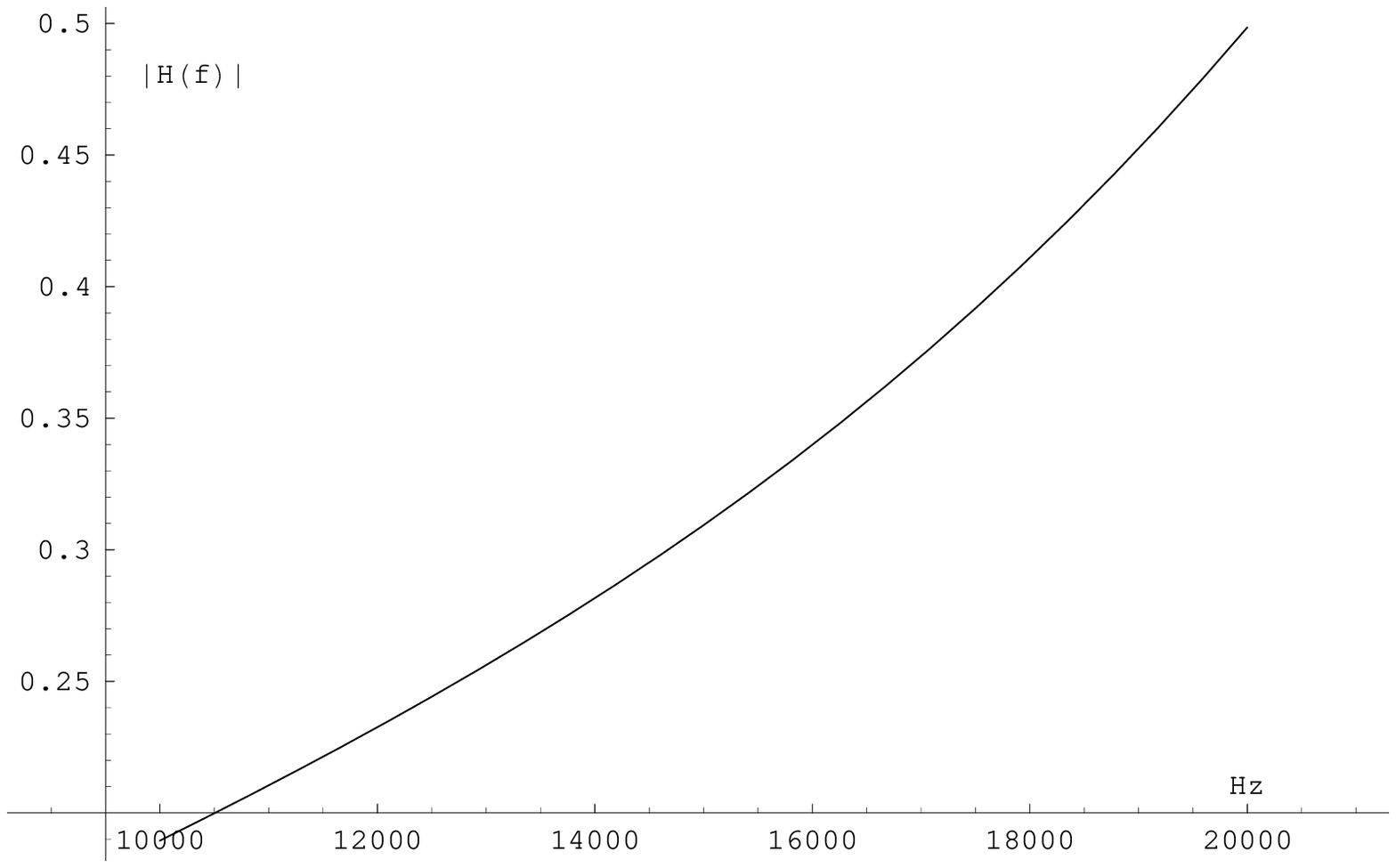}

\caption{the absolute value of the total response function of the Virgo interferometer
to the magnetic component of the $\times$ polarization of GWs for
$\theta=\frac{\pi}{4}$ and $\phi=\frac{\pi}{3}$ in the frequency-range
$10000Hz\leq f\leq20000Hz$}
\end{figure}
\begin{figure}
\includegraphics{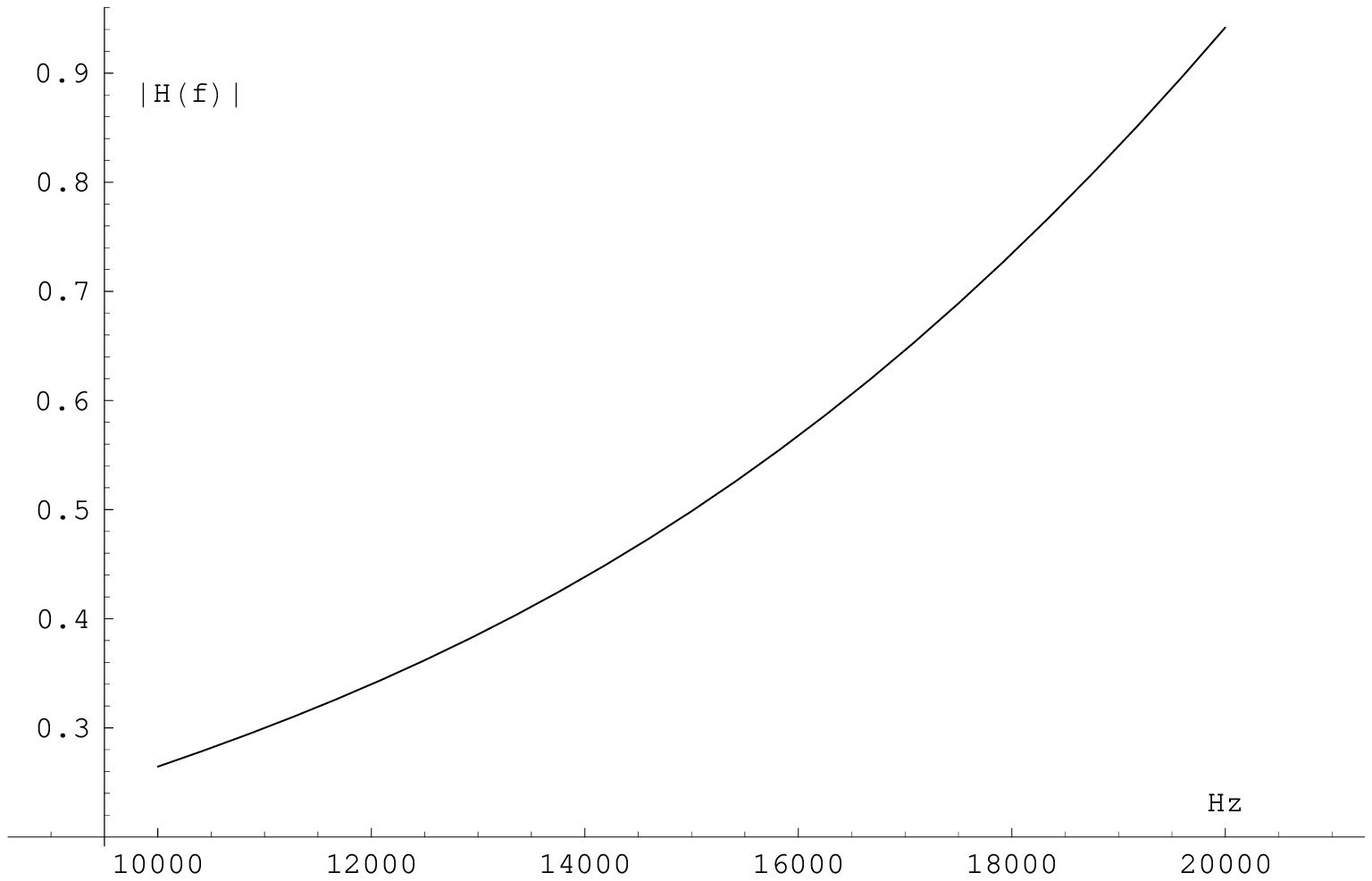}

\caption{the absolute value of the total response function of the LIGO interferometer
to the magnetic component of the $\times$ polarization of GWs for
$\theta=\frac{\pi}{4}$ and $\phi=\frac{\pi}{3}$ in the frequency-range
$10000Hz\leq f\leq20000Hz$}
\end{figure}

\section{Conclusion remarks}

The analysis of the response functions of interferometers for the
{}``magnetic'' components of GWs has been generalized in its full
frequency dependence, while in the work of Baskaran and Grishchuk
\cite{key-13} the response functions were computed only in the low
frequencies approximation (i.e. wavelength much larger than the linear
dimensions of the interferometer). It has also been shown that the
response functions to the magnetic components of GWs grows at high
frequencies, thus, because the value of the response functions to
the {}``electric'' components (i.e. the well known ordinary components)
decreases, the response functions to the magnetic components could
in principle begin the dominant part of the signal at high frequencies.
This could be important in the context of a detection of the signal
at high frequencies and confirms the result of Baskaran and Grishchuk
that the magnetic contributions must be taken into account in the
data analysis. We also emphasize that the fact that the response functions
of the magnetic components grow at high frequencies shows that, in
principle, the frequency-range of Earth-based interferometers could
extend to frequencies over 10000 Hz.

\section*{Acknowledgements}

I would like to thank Maria Felicia De Laurentis and Giancarlo Cella
for helpful advices during my work. The European Gravitational Observatory
(EGO) consortium has also to be thanked for the using of computing
facilities.

\end{document}